\documentclass[twocolumn,amsmath,showpacs,aps,prb,10pt,longbibliography]{revtex4-1}

\usepackage{array}
\usepackage{graphicx}
\usepackage{times}

\newcolumntype{C}[1]{>{\centering\arraybackslash}m{#1}}

\begin{document}

\title{Triangular lattice exciton model}

\author{Daniel~Gunlycke}
\email{daniel.gunlycke@nrl.navy.mil}
\address{Naval Research Laboratory, Washington, D.C. 20375, USA}
\author{Frank~Tseng}
\altaffiliation{National Research Council Research Associate}
\address{Naval Research Laboratory, Washington, D.C. 20375, USA}

\begin{abstract}
We present a minimalistic equilateral triangular lattice model, from which we derive electron and exciton band structures for semiconducting transition-metal dichalcogenides.  With explicit consideration of the exchange interaction, this model is appropriate across the spectrum from Wannier to Frenkel excitons.  The single-particle contributions are obtained from a nearest-neighbor tight-binding model parameterized using the effective mass and spin-orbit coupling.  The solutions to the characteristic equation, computed in direct space, are in qualitative agreement with first-principles calculations and highlight the inadequacy of the two-dimensional hydrogen model to describe the lowest-energy exciton bands.  The model confirms the lack of subshell degeneracy and shows that the A-B exciton split depends on the electrostatic environment as well as the spin-orbit interaction.
\end{abstract}


\maketitle


\section{Introduction}
\label{s.1}

It is well known that the electronic properties of a crystal depend critically on its crystal structure.  Yet, the lattice is usually absent from descriptions of low-energy electronic excitations in pristine semiconducting or insulating crystals.  The reason the lattice can often be ignored is that these excitations generate bound electron and hole pairs, known as excitons, that are typically either confined to single sites or span regions much larger than the relevant lattice constant.  The energies of the former site-confined excitons can be estimated by the energetics of the excited site and an ``exchange energy'' from neighboring sites.\cite{Frenkel31}  These excitons have been named after Frenkel and can be found in e.g. molecular solids.  The second type of excitons can be described by the hydrogen model adapted for excitons\cite{Wannier37} and are usually referred to as Wannier--Mott excitons and can be found in semiconductors with large permittivities and dispersive charge carriers.

Lattice effects could be important, however, in the description of excitons in semiconducting transition metal dichalcogenides such as molybdenum disulfide (MoS$_2$).  These lubricants, investigated in the 1960s,\cite{Frindt63,Evans65,Wilson69} have received renewed interest after the isolation of individual layers\cite{Novoselov05} and the demonstration that the monolayers, unlike their bulk counterpart, are direct gap\cite{Li07,Mak10} semiconductors.  An estimate for the separation of the electron and hole in a two-dimensional (2D) semiconductor is the ground state radial expectation value of the 2D hydrogen model\cite{Ralph65,Shinada66} for excitons $\langle r\rangle_{1s}\sim 0.09\varepsilon_r$\,nm, where $\varepsilon_r$ is the relative permittivity.  Relative permittivities are expected to be quite small in 2D materials,\cite{Molina11,Cheiwchanchamnangij12,Lin14} and for $\varepsilon_r\lesssim3.5$, the expectation value $\langle r\rangle_{1s}$ becomes smaller than the lattice constant $a$ in these materials, which necessitates atomistic treatment.  The exciton binding energy provides another rough measure of the degree to which the excitons are confined.  Unfortunately, there is no conclusive band edge feature in the optical spectroscopy data, which has resulted in different interpretations of observed spectroscopic features.\cite{Frindt63,Beal76,Chernikov14,Ye14,He14,Hanbicki15,Zhu15}  Furthermore, band gap predictions based on density functional theory depends significantly on calculation details.\cite{Cheiwchanchamnangij12,Kumar12,Ramasubramaniam12,Shi13,Li13,Liang13,Qiu13,Soklaski14,Liang14,Ye14}  Photoconductivity\cite{Frindt63,Klots14} and scanning tunneling spectroscopy\cite{Ugeda14} data suggests a binding energy of at least 0.5\,eV, providing further indication that the exciton radius is of the order of the lattice constant.

\begin{figure}
	\includegraphics{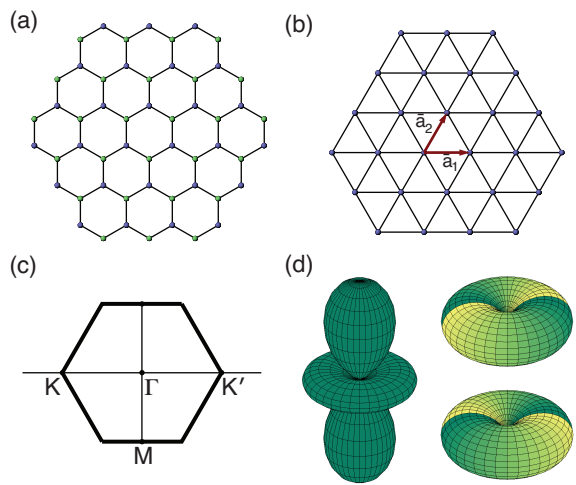}
	\caption{(Color online) (a) Top view of the hexagonal-tiled structure of trigonal prismatic monolayer transition metal dichalcogenides.  (b) The transition-metal sites form a triangular-tiled structure that exposes the underlying equilateral triangular lattice.  (c) First Brillouin zone of the equilateral triangular lattice with symmetry points $\Gamma$, K, K$'$, and M.  (d) The conduction and valence band d-orbitals at K and K$'$ have magnitudes that are azimuthally symmetric.}
	\label{f.1}
\end{figure}
We present a minimalistic exciton model, which we refer to as the triangular lattice exciton (3-ALE) model.  This model accounts for both the exchange and spin-orbit interactions and the equilateral triangular lattice shown in Fig.\,\ref{f.1}.  A dielectric constant or function is used to account for dielectric polarization.  Unlike typical exciton models solving the Bethe-Salpeter equation\cite{Fuchs08,Berghauser14,Konabe14} in reciprocal space, we derive a characteristic equation on sparse form in direct space, thus allowing great computational efficiency.  This efficiency enables large calculations and hence avoids the convergence challenge for tightly bound electron/hole pairs.\cite{Huser13}  The long-range Coulomb interaction in direct space causes no problem for tightly bound electron/hole pairs in transition-metal dichalcogenides.  Our model illustrates the breakdown of the 2D hydrogen model for low permittivity through the atomic scale electron-hole separation and the binding energies and oscillator strengths of the lowest exciton states.  In addition, the model shows that: (1) optically allowed exciton states depend sensitively on the shared-site potential, as well as the relative permittivity, (2) the degeneracy within lower shells is broken, and (3) the energy separation between the A and B excitons depends on the electrostatic environment, as well as the spin-orbit coupling.\cite{Frindt63}

The 3-ALE model can, as is usually the case for exciton models, be separated into electron and hole contributions and a contribution from their mutual Coulomb interaction.  Sections~\ref{s.2} and \ref{s.3} present the single-particle triangular lattice model and the complete 3-ALE model, respectively.  Section~\ref{s.4} provides conclusions about the 3-ALE model together with suggestions for possible generalizations.

\section{Triangular Lattice Model}
\label{s.2}

Monolayers of trigonal-prismatic transition-metal dichalcogenides have a hexagonal-tiled structure as shown in Fig.\,\ref{f.1}(a) with trigonal point group symmetry $D_{3h}$.  The bands around the Fermi level are transition metal $d$-bands\cite{Wilson69} with only small contributions from the chalcogenides entering through hybridization,\cite{Mattheiss73,Mattheiss73_2,Coehoorn87} which we will herein neglect.  This allows us to construct a basis comprised solely by transition-metal sites.  Without the chalcogenide sites, we have exposed the equilateral triangular lattice in these crystals shown in Fig.\,\ref{f.1}(b).  This triangular-tiled structure has a point group pseudosymmetry $D_{6h}$.  Pseudosymmetry can be useful, but it is imperative that one exercise caution as pseudosymmetry introduces artificial symmetry, in the present case inversion symmetry.

\begin{figure}[t]
	\includegraphics{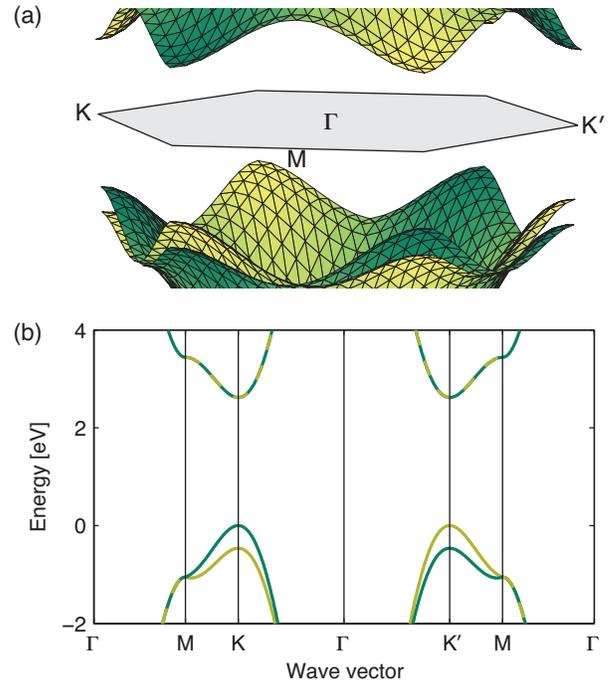}
	\caption{(Color online) (a) The electron band structure of the minimalistic lattice model captures the band gap, band curvature, and valence-band split at the K and K$'$ points.  (b) Additional parameters are needed for other band features such as a spin-split conduction band and a valence band with a local maximum at $\Gamma$ near its band edge.  Bright and dark curves indicate spin.  Parameters:  $E_g=2.6$\,eV, $t=0.89$\,eV, and $\Delta=0.463$\,eV.}
	\label{f.2}
\end{figure}
\begin{table}[b]
    \caption{Transition-metal dichalcogenide parameters provided in units of electron volts, electron mass, and nanometers.  The hopping parameter for each material obtained from the effective mass $m^*$ and lattice constant $a$ using $t\equiv2\hbar^2/3m^*a^2$.  $\Delta$ is the spin-orbit coupling.}
    \begin{tabular}{| C{0.8in} || C{0.55in} | C{0.55in} || C{0.55in} | C{0.55in} |}
        \hline
         & $\,m^*$ & $a$ & $t$ & $\Delta$ \\
        \hline
        \hline
        MoS$_2$ & 0.64 & 0.312 & 0.82 & 0.152 \\
        \hline
        MoSe$_2$ & 0.70 & 0.325 & 0.69 & 0.196 \\
        \hline
        WS$_2$ & 0.49 & 0.313 & 1.06 & 0.425 \\
        \hline
        WSe$_2$ & 0.54 & 0.325 & 0.89 & 0.463 \\
        \hline
    \end{tabular}
    \label{t.1}
\end{table}
\begin{figure*}
	\includegraphics{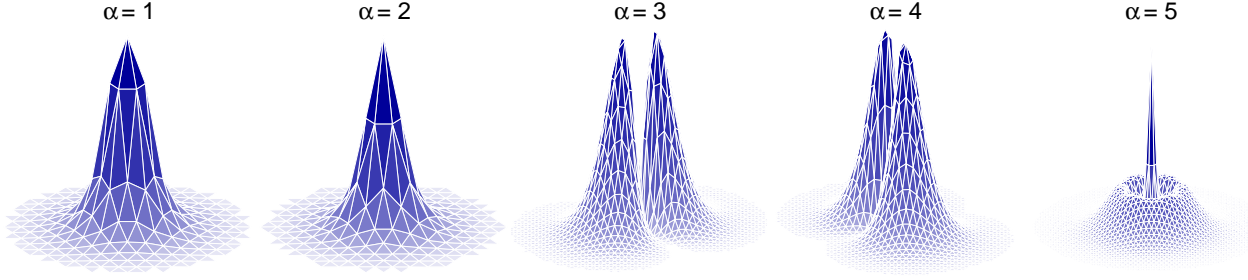}
	\caption{(Color online) Real-space probability density $|\xi_{\sigma\vec K\alpha\vec R}|^2$ of the electron for a centered hole in zone-center exciton states.  The triangular lattice grid shows that the electron and hole in the $\alpha=1$ ($1s$-A) and $\alpha=2$ ($1s$-B) exciton states are tightly bound and are likely to be found on the same or neighboring sites.  Parameters:  $t=0.89$\,eV, $\Delta=0.463$\,eV, $\varepsilon_r=3.2$ and $\Delta v_0=1.6$\,eV.}
	\label{f.3}
\end{figure*}
The triangular lattice is generated by a set of translation vectors $\vec R$, which are linear combinations of the lattice vectors $\vec a_1=a\hat x$ and $\vec a_2=a(\hat x+\sqrt{3}\hat y)/2$ with $a$ being the lattice constant.  We can introduce an electron field on the triangular lattice through the hermitian conjugate of the electron field operator
\begin{equation}
	\Psi(\vec r) = \sum_{n\sigma\vec R}c_{n\sigma\vec R}\,\phi_{n\sigma}(\vec{r}-\vec R),
	\label{e.1}
\end{equation}
where $c_{n\sigma\vec R}$ is the annihilation operator for an electron in the Wannier orbital\cite{Wannier37} $\phi_{n\sigma}(\vec r-\vec R)$ centered on site $\vec R$ with $n\in\{c,v\}$ and $\sigma\in\{\pm1/2\}$ being band and spin indices, respectively.  Within the tight-binding approximation, these Wannier orbitals can be viewed as isolated atomic orbitals.  The electronic states outside the band gap near the band edges are concentrated around the corners of the hexagonal Brillouin zone,\cite{Mak10} at the symmetry points K and K$'$, shown in Fig.\,\ref{f.1}(c).  These symmetry points are located at $\vec K_\tau=4\pi\tau\hat x/3a$ with $\tau\in\{\pm1\}$ being a valley index.  At these symmetry points, the $d$-orbitals in the conduction and valence bands take the form $d_{z^2}$ and $d_{x^2-y^2}\pm id_{xy}$, respectively, which are shown in Fig.\,\ref{f.1}(d).  These orbitals are generally hybridized\cite{Mattheiss73,Mattheiss73_2,Liu13} throughout the Brillouin zone.  This hybridization can be ignored, however, so long as we restrict the usage of our model to an energy regime near the band edges.  With this restriction in mind, we shall herein treat the the conduction and valence band as two non-interacting bands.  We also restrict the interaction between different localized Wannier orbitals to nearest neighbors, which relative to an arbitrary site are located at the six sites $\vec\delta\in\{\pm\vec a_1,\pm\vec a_2,\pm(\vec a_1-\vec a_2)\}$.  The Hamiltonian could then be expressed as $\hat{H}=\sum_{n\sigma}\hat{H}_{n\sigma}$ with
\begin{equation}
	\hat{H}_{n\sigma} = \sum_{\vec R\vec\delta}t_{n\sigma\vec\delta}c^\dagger_{n\sigma\vec R+\vec\delta}c_{n\sigma\vec R}+\sum_{\vec R}\varepsilon_nc^\dagger_{n\sigma\vec R}c_{n\sigma\vec R},
	\label{e.2}
\end{equation}
where $t_{n\sigma\vec\delta}$ and $\varepsilon_n$ are nearest-neighbor and onsite hopping parameters, respectively.  In the absence of the spin-orbit interaction, $t_{n\sigma\vec\delta}$ is spin-independent and isotropic.  The latter isotropy is evident from the lack of azimuthal dependence of the magnitude of the $d$-orbitals in Fig.\,\ref{f.1}(d).  To account for the spin-orbit interaction, which is significant in the transition metals, we add an imaginary term so that
\begin{equation}
	t_{n\sigma\vec\delta}=t_n+4i\sigma\tilde t_n\,\sin\vec K_+\cdot\vec\delta,
	\label{e.3}
\end{equation}
where $t_n$ and $\tilde t_n$ are real parameters.  
We have determined the introduced parameters above, up to an arbitrary reference energy, from the band gap $E_g$, the effective mass $m^*$, and the spin-orbit coupling $\Delta$:
\begin{equation}
	\begin{array}{lll}
		\varepsilon_c=3t+E_g \quad& t_c=t \quad& \tilde t_c=0,\\
		\varepsilon_v=-3t-\Delta/2 \quad& t_v=-t \quad& \tilde t_v=\Delta/18,
	\end{array}
	\label{e.4}
\end{equation}
where $t\equiv2\hbar^2/3m^*a^2$, and the band indices $c$ and $v$ refer to the the conduction and valence band, respectively.

The single-particle band structure
\begin{equation}
	\varepsilon_{n\sigma\vec k}=\varepsilon_n+\sum_{\vec\delta}t_{n\sigma\vec\delta}\,e^{-i\vec k\cdot\vec\delta},
	\label{e.5}
\end{equation}
is shown in Fig.\,\ref{f.2} for tungsten diselenide (WSe$_2$).  Parameters for additional semiconducting transition metal dichalcogenides are provided in Table~\ref{t.1}.

\section{Triangular Lattice Exciton Model}
\label{s.3}

When dealing with elementary excitations, it is convenient to perform a canonical transformation that takes into account electrons in the conduction band and holes in the valence band.  Consider an elementary excitation bringing an electron with spin $\sigma$ from a state in the valence band with wave vector $\vec k_v=\vec k-\vec K/2$ across the band gap into a state in the conduction band with wave vector $\vec k_c=\vec k+\vec K/2$.  This excitation creates an electron with $\vec k_e=\vec k_c$ and a hole with $\vec k_h=-\vec k_v$ so that the combined wave vector becomes $\vec K$.  This combined wave vector commutes with the Hamiltonian and is therefore a good quantum number.  Other quantum numbers are $\sigma$ and $\lambda$, the latter describing the relative motion of the electron and hole.  Defining the phase-modified annihilation operators
\begin{subequations}
\begin{alignat}{2}
	\tilde c_{\sigma\vec R}=c_{c\sigma\vec R}e^{-i\vec K\cdot\vec R/2},\\
	\tilde d_{\bar\sigma\vec R}=\sigma c^\dagger_{v\sigma\vec R}e^{-i\vec K\cdot\vec R/2},
\end{alignat}
	\label{e.6}%
\end{subequations}
 for an electron and hole, respectively, we can express a general excitation state as
\begin{equation}
	|\mathrm{exc}\rangle = \frac{1}{\sqrt{N}}\sum_{\vec R\vec R'}\xi_{\sigma\vec K\lambda\vec R}\tilde c^\dagger_{\sigma\vec R'+\vec R}\tilde d^\dagger_{\bar\sigma\vec R'}|0\rangle,
	\label{e.7}
\end{equation}
where $N$ is the number of lattice sites, $\xi_{\sigma\vec K\lambda\vec R}$ is the direct space eigenfunction for the electron/hole pair, and $|0\rangle$ is the vacuum state.  The focus below is to obtain $\xi_{\sigma\vec K\lambda\vec R}$ by solving for the relative motion of the electron and hole.

The Hamiltonian $\hat{H}$ describing the electron-hole pair contains a mutual Coulomb interaction as well as the single-particle contributions described above.  We neglect Coulomb contributions from electrons in the valence bands and approximate the screened Coulomb integrals as
\begin{equation}
	V_{\vec R} \equiv \left\{
	\begin{array}{cc}
		\frac{e^2}{4\pi\varepsilon_r\varepsilon_0|\vec R|} & \qquad(\vec{R}\ne 0), \\
		\Delta v_0 & \qquad(\vec{R} = 0),
	\end{array}
	\right.
	\label{e.8}
\end{equation}
where $e$ is the elementary charge, $\varepsilon_0$ is the vacuum permittivity, and $\Delta v_0\equiv v_0-w_0$ is the difference between the direct integral $v_0$, and the exchange integral $w_0$, the latter assumed to be negligible when the electron and hole occupy different sites.  

In analogy with the hydrogen atom, we choose the origin to be position of the positive charge carrier, in this case, the hole.  We also let this origin track the motion of the hole, thereby transfering the kinetic energy of the hole over to the electron.  This allows us to express the hole-centered electron/hole pair Hamiltonian as
\begin{equation}
	\hat{H}_{\sigma\vec K} = \sum_{\vec R\vec\delta}T_{\sigma\vec K\vec\delta}\,\tilde c^\dagger_{\sigma\vec R+\vec\delta}\tilde c_{\sigma\vec R}+\sum_{\vec R}(E_0-V_{\vec R})\tilde c^\dagger_{\sigma\vec R}\tilde c_{\sigma\vec R},
	\label{e.9}
\end{equation}
where the coefficients
\begin{subequations}
\begin{alignat}{2}
	T_{\sigma\vec K\vec\delta}&\equiv t_{c\sigma\vec\delta}\,e^{-i\vec K\cdot\vec\delta/2}-t_{v\sigma\vec\delta}\,e^{i\vec K\cdot\vec\delta/2},\\
	E_0&\equiv\varepsilon_c-\varepsilon_v-E_g,
\end{alignat}
	\label{e.10}%
\end{subequations}
can be obtained from Eqs. (\ref{e.3}) and (\ref{e.4}) with parameters from Table \ref{t.1}.
\begin{table}[b]
    \caption{Hydrogen shell labels for the lowest five quantum numbers describing excitations from the valence band (Series A) and split-off band (Series B) resulting from the spin-orbit coupling.}
    \begin{tabular}{| C{1.1in} || C{0.38in} | C{0.38in} | C{0.38in} | C{0.38in} | C{0.38in} |}
        \hline
        $\mathrm{Quantum~number}$ & $\alpha=1$ & $\alpha=2$ & $\alpha=3$ & $\alpha=4$ & $\alpha=5$ \\
        \hline\hline
        $\mathrm{Series~A~label}$ & $1s$ & & $\,2p_x$ & $\,2p_y$ & $2s$ \\
        \hline
        $\mathrm{Series~B~label}$ & & $1s$ & & & \\
        \hline
    \end{tabular}
    \label{t.2}
\end{table}
\begin{figure}
	\includegraphics{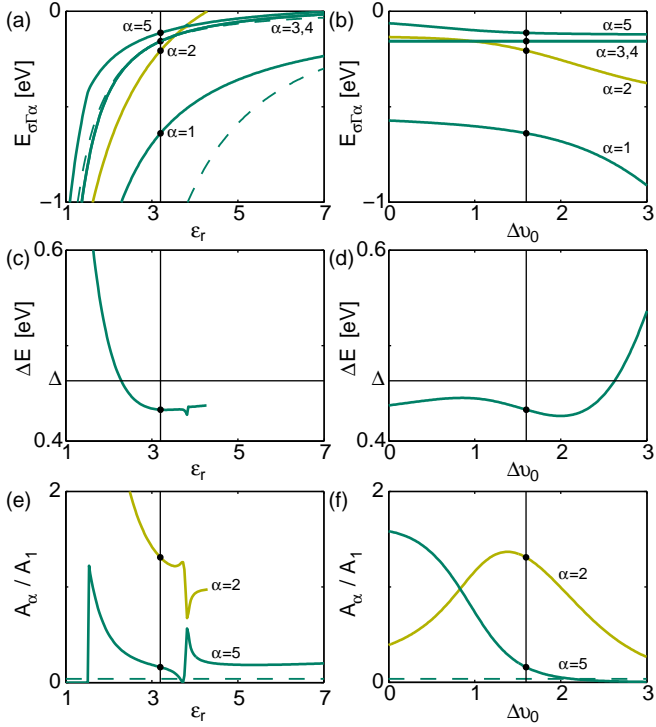}
	\caption{(Color online) Zone-center excitons.  Exciton energy as a function of the relative permittivity (a) and onsite Coulomb interaction (b) with the intersection $\varepsilon_r=3.2$ and $\Delta v_0=1.6$ indicated by vertical lines.  Solid and dashed curves obtained from the 3-ALE model and the the 2D hydrogen model, respectively.  While there is good agreement for the $\alpha=3,4$ ($2p$) curve, the latter model breaks down for the tightly bound electron/hole pairs in the $\alpha=1$ ($1s$--A) and $\alpha=2$ ($1s$--B) states.  The energy separation between these states depends on the relative permittivity (c) and onsite Coulomb interaction (d), as well as the spin-orbit coupling $\Delta$.  The oscillator strengths relative to that of $\alpha=1$ ($1s$--A) are shown in (e) and (f).  Dark and bright curves represent exciton states from the A series and B series, respectively.  Additional parameters:  $t=0.89$\,eV and $\Delta=0.463$\,eV.}
	\label{f.4}
\end{figure}

The solutions of the characteristic equation for the relative motion
\begin{equation}
	\hat{H}_{\sigma\vec K}|\sigma\vec K\lambda\rangle=E_{\sigma\vec K\lambda}|\sigma\vec K\lambda\rangle
	\label{e.11}
\end{equation}
provide the electron/hole pair eigenenergies $E_{\sigma\vec K\lambda}$ and eigenstates of the form
\begin{equation}
	|\sigma\vec K\lambda\rangle = \sum_{\vec R}\xi_{\sigma\vec K\lambda\vec R}\tilde c^\dagger_{\sigma\vec R}|0\rangle.
	\label{e.12}
\end{equation}
These states can be divided into two sets:  (1) bound electron/hole-pair states known as excitons and (2) unbound electron/hole-pair states.  The former states appear at discrete energies and are described by $\lambda=\alpha$, where we let $\alpha$ take positive integers counted from the strongest bound state.  The latter unbound states are continuous in energy and are described by $\lambda=\vec k$.

The exciton states with the wave vector $\vec K$ at the zone center $\Gamma$ are of particular importance for optoelectronic experiments due to the relative high speed of light causing almost vertical electronic excitations.  The eigenfunctions of the five strongest bound $\Gamma$ excitons are shown in Fig.\,\ref{f.3}, where each vertex represents a transition metal lattice site.  Strictly, the lattice symmetry prohibits the exciton quantum numbers $\alpha$ from being separated into radial and azimuthal quantum numbers.  For the lowest-energy bands, however, designation using shell labels from the hydrogen model still makes sense and we provide the conversion in Table \ref{t.2}.

The exciton energies depend on the screened Coulomb interaction, as depicted in Fig.\,\ref{f.4}.  First, we note that the $\alpha=3,4$ ($2p$) curves are lower in energy than the $\alpha=5$ ($2s$) curve, in agreement with first-principles predictions.\cite{Ye14}  Second, we note that the $\alpha=2$ ($1s$--B) exciton only appears below the band edge for a sufficiently small dielectric constant.  In this regime, the hydrogen model also breaks down for the A exciton series due to the tightly bound electron/hole pairs.  Third, the $\alpha=3,4$ ($2p$) states have a node at the origin and are unaffected by $\Delta v_0$.  Fourth, the energy separation between the $\alpha=1$ and $\alpha=2$ states (``the well-known A-B exciton split'') is not only a function of the spin-orbit coupling but also the Coulomb interaction.  Fifth, the Coulomb interaction also affects the relative oscillator strengths $A_\alpha/A_1\equiv|\xi_{\sigma\vec 0\alpha\vec 0}|^2/|\xi_{\sigma\vec 01\vec 0}|^2$ and could be modulated through the dielectric environment.

\begin{figure}
	\includegraphics{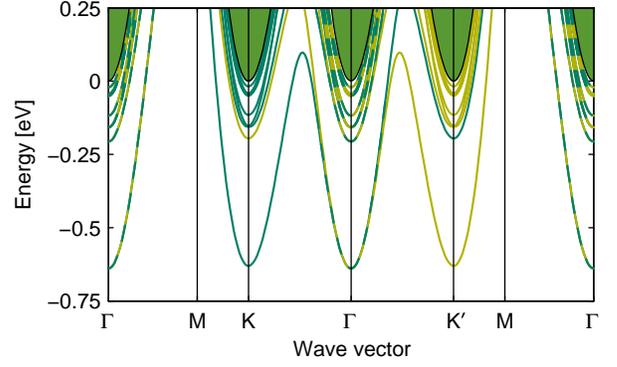}
	\caption{(Color online) Electron/hole pair band structure.  Discrete bands and filled regions contain bound exciton states and free electron/hole pairs, respectively.  Bright and dark bands indicate spin of the excited electron.  The bands, enumerated at $\Gamma$ starting from the strongest bound state, are: $\alpha=1$ ($1s$--A), $\alpha=2$ ($1s$--B), $\alpha=3,4$ ($2p$), and $\alpha=5$ ($2s$).  Parameters:  $t=0.89$\,eV, $\Delta=0.463$\,eV, $\varepsilon_r=3.2$, and $\Delta v_0=1.6$\,eV.}
	\label{f.5}
\end{figure}
Exciton states away from the zone center could also potentially be exploited in applications.  The energies of these states, $E_{\sigma\vec K\alpha}$, are shown as discrete bands in Fig.\,\ref{f.5}.  Reflection symmetry or time-reversal symmetry causes the zone-center states to be doubly degenerate and requires the K excitons to have the same energies as the K$'$ excitons with the opposite excited spin.\cite{Yao08,Xiao12}  This degeneracy could be broken, however, by an external magnetic field.\cite{Li14,MacNeill15}  Transitions between exciton states are possible through additional interactions, as long as they exhibit appropriate symmetry.\cite{Song13,Glazov14}  First-principles band structures indicate that there might be additional states near K and K$'$ originating from the valence band $\Gamma$ point.  To capture these states, the single-particle basis would need to be expanded.

The electron/hole pair band structure also contains unbound states.  Unbound states have been derived analytically for the 2D hydrogen model.\cite{Shinada66}  Obtaining similar solutions using the 3-ALE model is not practical, owing to the large number of sites needed for such calculations.  Therefore, we model instead the filled regions in Fig.\,\ref{f.5} using free electron/hole pairs.  For these free electron/hole pairs, the Coulomb interaction $v_{\vec R}\rightarrow0$.  The eigenfunctions are then Bloch waves given by  $\xi_{\sigma\vec K\vec k\vec R}=N^{-1/2}\,e^{i\vec{k}\cdot{\vec R}}$.  The corresponding energies of the free electron/hole pairs are
\begin{equation}
	E_{\sigma\vec K\vec k} = E_0+\sum_{\vec\delta}t_{\sigma\vec K\vec\delta}\,e^{-i\vec k\cdot\vec\delta},
	\label{e.13}
\end{equation}
where we have as expected $E_{\bar\sigma\vec K\vec k}=E_{\sigma-\vec K-\vec k}$, again from time-reversal symmetry.

\section{Conclusions}
\label{s.4}

The model presented herein have been derived from the many-body Hamiltonian with the goal to extract the dominant physics among the elementary excitations in 2D semiconductors with a equilateral triangular lattice.  The approximations should be reasonable for tightly bound electron/hole pairs, and we therefore expect the model to work well for optically driven transitions into the lower exciton states.  The predicted results are in qualitative agreement with first-principles quasi-particle calculations, and for improved accuracy, we suggest using this model with a more detailed dielectric environment with a complex dielectric function that is spatially\cite{Berkelbach13} and/or energy dependent.\cite{Li14_,Mukherjee15}  To achieve a better description away from the band edges, which could be needed for zone-corner excitons, it is also recommended to use a larger basis to account for $d$-band hybridization.\cite{Zahid13,Cappelluti13,Liu13}  During the final stage of this paper, the authors have become aware of a recent paper including such hybridization effects for excitons in MoS$_2$.\cite{Wu15}

In summary, we have presented a minimalistic and computationally efficient model capturing the essential physics of low-energy electronic excitations.  This model could be used to understand experimental absorption spectra and is well suited to serve as a foundation for more advanced models, which could describe exciton coupling to, e.g., defects, interfaces, vibrations, and externally applied fields.


\begin{acknowledgments}
This work has been supported by the Office of Naval Research (ONR), directly and through the Naval Research Laboratory (NRL).  The authors thank Brett Dunlap, Ergun Simsek, and Carter White for discussions.  F.\,T. acknowledges support from NRL through the National Research Council (NRC) Research Associateship Program.
\end{acknowledgments}


%


\end{document}